\pdfoutput=1

\documentclass[11pt]{article}

\usepackage{emnlp2021}

\usepackage{flushend}
\usepackage{times}
\usepackage{latexsym}
\usepackage{microtype}
\usepackage{graphicx}
\usepackage{booktabs}  
\usepackage{multirow}
\usepackage{subfigure}
\usepackage{verbatim}
\usepackage{CJKutf8} 
\usepackage{amssymb}
\usepackage{amsmath}

\usepackage{algorithm}
\usepackage{algorithmic}
\usepackage{makecell}
\usepackage[T1]{fontenc}

\usepackage[utf8]{inputenc}

\usepackage{microtype}

\title{Uni-FedRec: A Unified Privacy-Preserving News Recommendation\\  Framework for Model Training and Online Serving}

\author{Tao Qi$^1$, Fangzhao Wu$^2$, Chuhan Wu$^1$, Yongfeng Huang$^1$ and Xing Xie$^2$\\
  $^1$Department of Electronic Engineering \& BNRist, Tsinghua University, Beijing 100084, China  \\
  $^2$Microsoft Research Asia, Beijing 100080, China\\
  {\tt \{taoqi.qt, wufangzhao, wuchuhan15\}@gmail.com}\\
  {\tt yfhuang@tsinghua.edu.cn}\\
  {\tt xing.xie@microsoft.com}
  }

\begin{document}
\maketitle
\begin{abstract}

News recommendation is important for personalized online news services.
Most existing news recommendation methods rely on centrally stored user behavior data to both train models offline and provide online recommendation services.
However, user data is usually highly privacy-sensitive, and centrally storing them may raise privacy concerns and risks.
In this paper, we propose a unified news recommendation framework, which can utilize user data locally stored in user clients to train models and serve users in a privacy-preserving way.
Following a widely used paradigm in real-world recommender systems, our framework contains two stages.
The first one is for candidate news generation (i.e., recall) and the second one is for candidate news ranking (i.e., ranking).
At the recall stage, each client locally learns multiple interest representations from clicked news to comprehensively model user interests.
These representations are uploaded to the server to recall candidate news from a large news pool, which are further distributed to the user client at the ranking stage for personalized news display.
In addition, we propose an interest decomposer-aggregator method with perturbation noise to better protect private user information encoded in user interest representations.
Besides, we collaboratively train both recall and ranking models on the data decentralized in a large number of user clients in a privacy-preserving way.
Experiments on two real-world news datasets show that our method can outperform baseline methods and effectively protect user privacy.

%

\end{abstract}

\section{Introduction}

Online news platforms usually rely on personalized news recommendation techniques to help users obtain their interested news information~\cite{qi2021pprec,wu2019neurald}.
Existing news recommendation models usually exploit users' historical behavior data to model user interests for matching candidate news~\cite{wang2020fine,wu2019neurald,wutanr,wuuser,qi2021kim,ge2020graph,wu2021uag}.
For example, \citet{okura2017embedding} employed a GRU network to build user embeddings from browsed news.
\citet{wu2019ijcai} employed an attention network to build user embeddings by aggregating different clicked news.
Both of them match candidate news and user interests via the inner product of their embeddings.
In short, most of these methods rely on centralized storage of user behavior data to train models and serve users.
However, user behavior data is usually highly privacy-sensitive~\cite{chai2019secure}, and centrally storing them may arouse users' concerns on privacy leakage and violate some privacy protection regulations such as GDPR\footnote{https://gdpr-info.eu/}.

A few methods explore to recommend news in a privacy-preserving way~\cite{qi2020privacy}.
For instance, \citet{qi2020privacy} proposed to store user data in user clients and applied federated learning technique~\cite{mcmahan2017communication} to train news recommendation models on decentralized data.
In general, these methods usually focus on developing privacy-preserving model training approaches based on decentralized user behavior data for ranking candidate news.
However, how to generate candidate news and serve users in a privacy-preserving way remains an open problem.

In this paper, we propose a unified news recommendation framework based on federated learning (named \textit{Uni-FedRec}), which can utilize user behavior data locally stored in user clients to train models offline and serve users online in a privacy-preserving way.
Following a widely applied paradigm in real-world recommender systems~\cite{wu2021two,pal2020pinnersage}, \textit{Uni-FedRec} contains a recall stage for personalized candidate news generation and a ranking stage for candidate news ranking.
In the recall stage, the user client first locally learns multiple interest representations from clicked news to model diverse user interests.
These representations are further uploaded to the server to recall a small number of candidate news (e.g., 100) from a large news pool.
In the ranking stage, recalled candidate news are distributed to the user client and locally ranked for news personalized display.
Bedsides, user interest representations may encode user privacy information.~\cite{wu2020fedctr}.
To protect private user information encoded in interest representations, we propose an interest decomposer-aggregator method with perturbation noise to synthesize interest representations with a group of basic interest embeddings.
In addition, \textit{Uni-FedRec} utilizes user data decentralized in a large number of user clients to collaboratively train the recall and ranking model in a privacy-preserving way.
Extensive experiments on two real-world datasets verify that our method can significantly outperform baseline methods and effectively protect user privacy.

In summary, our contributions are as follows:

\begin{itemize}
    \item We propose a unified privacy-persevering news recommendation framework which can train model offline and serve users online with locally stored user data.

    \item We propose a privacy-preserving recall model which can comprehensively model diverse user interests and protect user privacy.
    
    \item Extensive experiments demonstrate that our framework can outperform many baseline methods and effectively protect user privacy.
\end{itemize}




\section{Related Work}

\subsection{Personalized News Recommendation}

Personalized news recommendation is an important research problem and has been widely studied in recent years~\cite{konstan1997grouplens,wang2011collaborative,liu2010personalized,bansal2015content,wu2020sentirec,qi2021hierec,wu2020ptum,wu2021empowering,wang2020fine,ge2020graph,an2019neural}.
Existing news recommendation methods aim to match candidate news content with user preferences mined from users' historical behaviors~\cite{khattar2018weave,wu2021personalized,wu2020fairness,ge2020graph,qi2021kim,wu2019neuralc,an2019neural}.
For example, \citet{okura2017embedding} proposed to learn user interest embeddings from the sequential information of user's clicked news via a GRU network.
\citet{an2019neural} proposed to model short-term user interest from news clicks via a GRU network and model long-term user interest via user ID embeddings.
They further combine them to form a unified interest embedding. 
\citet{wu2019neuralc} employed a multi-head self-attention network to learn user interest embeddings by modeling relatedness of users' reading behaviors.
Besides, all of these three methods performed the matching between candidate news and user interest via the inner product of their embeddings.
In brief, most of these methods rely on the centralized storage of user behavior data to train models and serve users.
However, users' behavior data is usually highly privacy-sensitive, and storing them in the server may arouse risks and user concerns on privacy leakage, and may also violate some privacy protection regulations (e.g., GDPR)~\cite{muhammad2020fedfast,wu2020fedctr}.
Different from these methods, we propose a unified privacy-preserving framework for news recommendation, which can utilize decentralized user behavior data to train models and serve users.

\subsection{Privacy-Preserving Recommendation}

\begin{figure*}
    \centering
    \resizebox{0.99\textwidth}{!}{
    \includegraphics{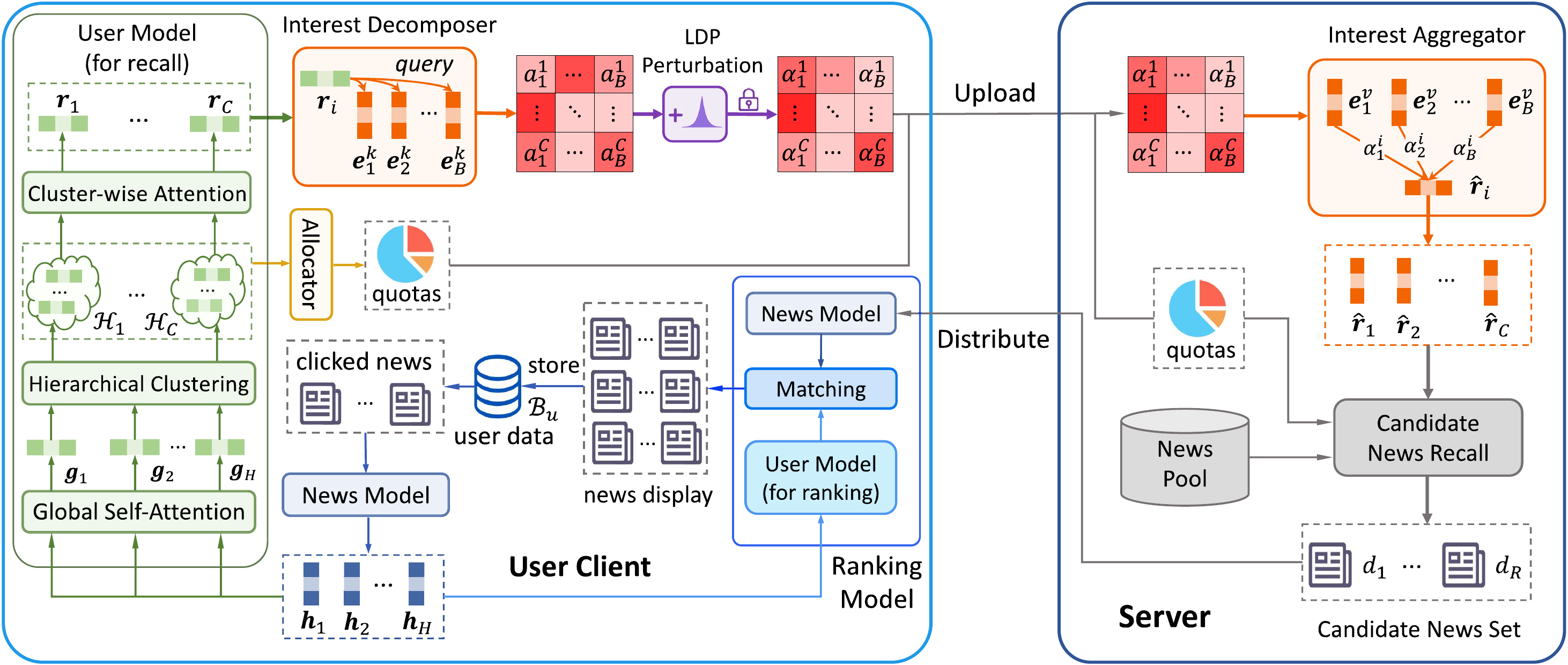}
    }
    \caption{The framework of \textit{Uni-FedRec} for privacy-preserving online serving.}
    \label{fig.serve}
\end{figure*}

Recently, due to users' increasing concerns on privacy leakage, some privacy-preserving recommendation methods have been proposed~\cite{qi2020privacy,flanagan2020federated,lin2020fedrec,wang2021fast,muhammad2020fedfast,yang2021fcmf,wu2021fedgnn,wu2020fedctr}.
For example, \citet{chai2019secure} proposed to compute gradients of user and item embeddings in user clients based on locally stored user rating data and upload gradients to the server for federated model updating.
Besides, to better protect user privacy, they employed the homomorphic encryption technique~\cite{gentry2009fully} to encrypt the uploaded gradients.
\citet{qi2020privacy} proposed to apply federated learning technique to train neural news recommendation models on decentralized user data.
They used local differential privacy technique~\cite{ren2018textsf} to protect the uploaded gradients from leaking user privacy.
In brief, most of these methods focus on training a recommendation model for ranking candidate news in a privacy-preserving way.
However, how to generate candidate news from news pool according to user interest and serve users with decentralized user behavior data are still unsolved problems.
Different from these methods, we propose a unified privacy-preserving news recommendation framework, which can utilize locally stored user data to generate candidate news from the server, and further serve users via local candidate news ranking. 

\section{Uni-FedRec}

In this section, we will introduce our unified privacy-preserving news recommendation framework (named \textit{Uni-FedRec}), which can utilize decentralized user data to train models and serve users.

\subsection{Framework Overview}

In \textit{Uni-FedRec}, user behavior data (e.g., displayed news and clicked news) is locally stored in user clients, and the news pool is stored and maintained in the server.
Following a widely used paradigm in real-world recommender systems~\cite{pal2020pinnersage,liu2020octopus}, \textit{Uni-FedRec} contains a recall stage for candidate news generation and a ranking stage for candidate news ranking.
To serve a user, the user client first employs a privacy-preserving recall model to locally learn multiple interest representations from clicked news to model diverse user interests.
The interest representations are further uploaded to the server to recall candidate news from a large news pool.
In the ranking stage, recalled candidate news are distributed to the user client and locally ranked for personalized news display.
To train models on decentralized user data, \textit{Uni-FedRec} coordinates massive user clients to collaboratively calculate gradients from their local user data for federated model updating.
Next, we will introduce each module of \textit{Uni-FedRec} in detail.

\subsection{Privacy-Preserving Recall Model}

As shown in Fig.\ref{fig.serve}, our privacy-preserving recall model contains four major modules, i.e., a \textit{user model}, an \textit{interest decomposer}, an \textit{LDP perturbation module} and an \textit{interest aggregator}.
The former is used to learn multiple interest representations to model diverse user interests.
The latter three are used to protect users' private information encoded in interest representations.

\begin{figure*}
    \centering
    \resizebox{0.7\textwidth}{!}{
    \includegraphics{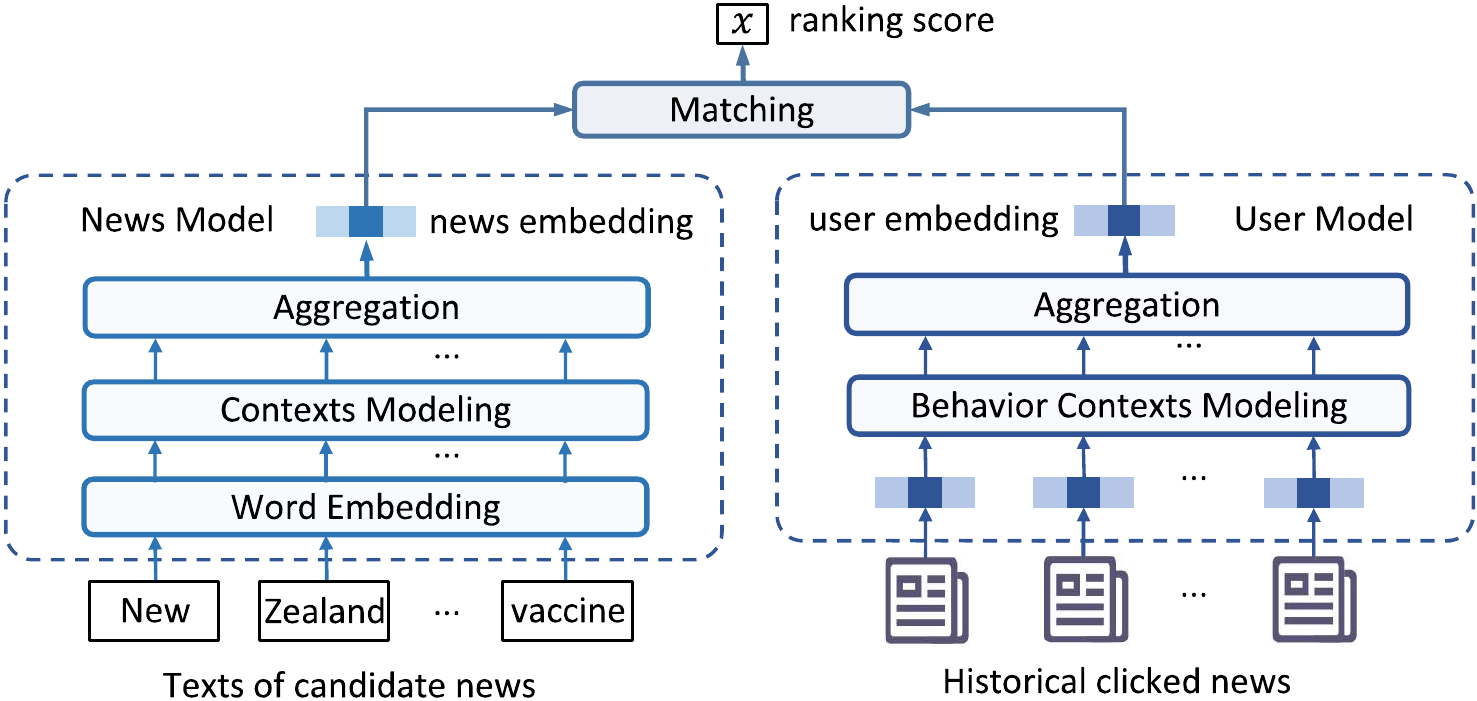}
    }
    \caption{Framework of the news ranking model.}
    \label{fig.rank_model}
\end{figure*}

\textbf{Interest Modeling:}
User behavior contexts are informative for modeling user interests~\cite{wu2019neuralc}.
In \textit{user model}, we first use a global self-attention network~\cite{vaswani2017attention} to learn context-sensitive representations of clicked news $[\textbf{g}_1,...,\textbf{g}_H]$ from representations of clicked news $[\textbf{h}_1,...,\textbf{h}_H]$, where $H$ is the number of clicked news\footnote{We introduced how to learn news representations from news texts in Section~\ref{sec.rank_model}.}.
Besides, users usually have diverse interests in multiple interest fields~\cite{pal2020pinnersage,liu2020octopus}. 
To capture diverse user interests, we divide user's clicked news into different interest clusters $\{\mathcal{H}_i|i=1,..., C\}$ via the hierarchical clustering algorithm~\cite{johnson1967hierarchical}, where $\mathcal{H}_i$ is the $i$-th cluster, and $C$ is the number of clusters.
The algorithm hierarchically merges the clusters until the average distances between any two clusters are larger than a given threshold $d_c$.
Then, we apply a cluster-wise attention network to learn unified interest representation $\textbf{r}_i$ for each cluster $\mathcal{H}_i$:
\begin{equation}
    \textbf{r}_i = \sum_{j=1}^{|\mathcal{H}_i|} \gamma^i_j\textbf{g}^i_j,\  \gamma^i_j = \frac{\exp(Att(\textbf{g}^i_j))}{\sum_{k=1}^{|\mathcal{H}_i|} \exp(Att(\textbf{g}^i_k))},
\end{equation}
where $\gamma^i_j$ and $\textbf{g}^i_j$ is the attention weight and representation for the $j$-th clicked news in the $i$-th cluster, and $Att(\cdot)$ is a dense network for calculating attention scores.
In this way, we can obtain multiple interest representations $\{\textbf{r}_i|i=1,...,C\}$ to model user interests in different interest fields.

\textbf{Privacy Protection:}
User interest representations may contain some private user information~\cite{wu2020fedctr}.
To protect private information encoded in interest representations, we propose an interest aggregator-decomposer method with permutation noise.
Its core is to synthesize interest representations by combining a group of trainable basic interest embeddings $\{(\textbf{e}^k_i,\textbf{e}^v_i)|i=1,...,B\}$ (\textit{BIE}) shared among different users, where $\textbf{e}^k_i$ and $\textbf{e}^v_i$ is the key and value of the $i$-th basic interest embedding, respectively, and $B$ is the number of \textit{BIE}.
In \textit{interest decomposer}, we first decompose each interest representation $\textbf{r}_i$ on the keys of \textit{BIE}: $a^i_j = Query(\textbf{r}_i,\textbf{e}^k_j)$, where $Query(\textbf{x},\textbf{y})$ is a query function implemented by dot product of $\textbf{x}$ and $\textbf{y}$, $a^i_j \in \mathbb{R}$ is the decomposition score of decomposing $\textbf{r}_i$ on embedding $\textbf{e}^k_j$.
We further perturb decomposition scores via local differential privacy (LDP) technique:
\begin{equation}
    \hat{a}^i_j = f_\delta (a^i_j) + n_I,\quad n_I\sim La(0,\lambda_I),
\end{equation}
where $\hat{a}^i_j$ is the protected decomposition score, $f_\delta(z)$ is a function for clipping $z$ with the scale of $\delta$, $n_I$ is a zero-mean Laplace noise, and $\lambda_I$ is its intensity.
Next, in the \textit{interest aggregator}, we further synthesize protected interest representations by combining value embeddings of \textit{BIE}:
\begin{equation}
    \hat{\textbf{r}}_i = \sum_{j=1}^B \alpha^i_j \textbf{e}^v_j, \quad \alpha^i_j = \frac{\exp( \hat{a}^i_j)}{\sum_{k=1}^B \exp( \hat{a}^i_k)},
\end{equation}
where $\hat{\textbf{r}}_i$ is the protected representation for $\textbf{r}_i$.

\textbf{News Recall:}
We further use each protected interest representation $\hat{\textbf{r}}_i$ to recall top $R_i$ candidate news that has the largest relevance with $\hat{\textbf{r}}_i$ from news in the pool.
We use inner product similarity to measure representation relevance, and this recall progress can be speeded up by some search algorithms such as ANN search~\cite{arya1998optimal}.
Besides, we have an \textit{allocator} to allocate quotas, i.e., the number of candidate news recalled by each channel.
We utilize ratios of clicked news belonging to each interest channel to generate their quotas:
\begin{equation}
    R_i = q_i \times R, \quad q_i = \frac{|\mathcal{H}_i|}{\sum_{j=1}^C |\mathcal{H}_j| }
\end{equation}
where $R$ is the total number of recalled candidate news.
Finally, we integrated candidate news of different channels and obtain the candidate news set $\mathcal{R} = \{d_i|i=1,...,R\}$, where $d_i$ is the $i$-th recalled candidate news.

\textbf{Loss Function:}
InfoNCE loss~\cite{oord2018representation,wu2019neuralc,an2019neural} is usually used to formulate loss function in recommendation task.
It requires a unified score to rank positive and negative samples, while our method will generate $C$ different recall scores for each news.
To tackle this issue, we combine recall scores of news $d$ generated by each channel $\hat{z}_i$ to form a unified score: 
\begin{equation}
  z = \sum_{i=1}^C q_i \hat{z}_i, \quad  \hat{z}_i=\textbf{d}\cdot \hat{\textbf{r}}_i,
\end{equation}
where $\textbf{d}$ is representation of $d$.
Next, for each positive sample, we randomly select $K_r$ negative samples from all news that are displayed to but not clicked by this user.
Then, we obtain the loss function $\mathcal{L}^r_u$ based on behavior data $\mathcal{B}_u$ of user $u$:
\begin{equation}
    \mathcal{L}^r_u = \sum_{i=1}^{|\mathcal{B}_u|} \frac{\exp(z_i)}{\exp(z_i) + \sum_{j=1}^{K_r}\exp(z^j_i)},
    \label{eq.loss}
\end{equation}
where $z_i$ and $z_i^j$ is the unified score of the $i$-th positive sample and its $j$-th negative sample.

\begin{figure*}
    \centering
    \resizebox{0.99\textwidth}{!}{
    \includegraphics{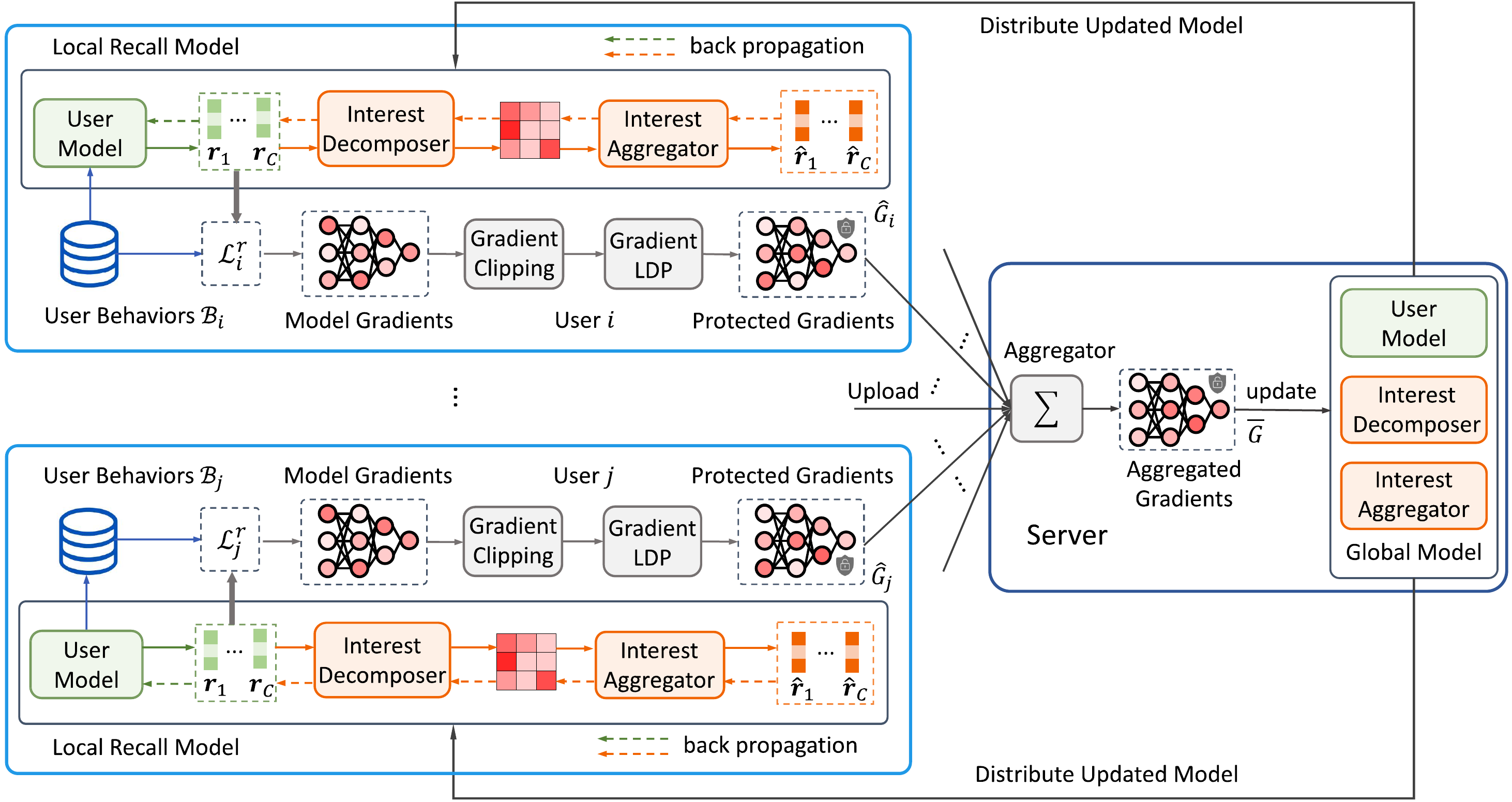}
    }
    \caption{The framework of \textit{Uni-FedRec} for privacy-preserving model training.}
    \label{fig.training}
\end{figure*}

\subsection{Ranking Model}
\label{sec.rank_model}
\textit{Uni-FedRec} contains a ranking model to locally rank candidate news in the user client.
Since local news ranking will not leak user privacy, we directly employ existing news ranking methods such as \textit{NRMS}~\cite{wu2019neuralc} in \textit{Uni-FedRec}.
As shown in Fig.~\ref{fig.rank_model}, these methods share a similar framework, where a news model learns news embedding from news texts, a user model learns user embedding from clicked news, and a matching module (e.g., dot product) matches candidate news and user interests for personalized ranking.\footnote{We utilize news model in the ranking model to generate news representation for the recall model.}
The news model is usually based on the stack of a word embedding layer, a context modeling layer (e.g., Transformer) and an aggregation layer (e.g., attention network), and the user model is usually based on the stack of a behavior context modeling layer (e.g., GRU) and an aggregation layer (e.g., attention network).
We also formulate the loss function $\mathcal{L}^g_u$ of the ranking model via the infoNCE loss:
\begin{equation}
    \mathcal{L}^g_u = \frac{1}{|\mathcal{B}_u|}\sum_{i=1}^{|\mathcal{B}_u|} \frac{\exp(x_i)}{\exp(x_i) + \sum_{j=1}^{K_g} \exp(x_i^j)},
\end{equation}
where $x_i$ and $x_i^j$ is the ranking score of the $i$-th positive sample and its $j$-th negative sample randomly selected from the same news impression respectively and $K_g$ is the number of negative samples.

\subsection{Privacy-Preserving Online Serving}

In Fig.~\ref{fig.serve}, we show the privacy-preserving framework of \textit{Uni-FedRec} for online serving with decentralized user data.
In \textit{Uni-FedRec}, users' behavior data is locally stored in user clients and is never uploaded to the server, which can effectively alleviate users' privacy concerns.
The server stores and maintains news articles in a news pool.
Besides, both the user client and the server contain the whole recall model.
When a user visits the news platform, the client first employs the recall model (i.e., \textit{user model}, \textit{interest decomposer} and \textit{perturbation module}) to build protected decomposition weights $\alpha^i_j$ and generate quotas of interest channels, which are further uploaded to the server.\footnote{We upload protected decomposition weights instead of interest representations to solve communication costs.}
After receiving them, the server further employs the \textit{interest aggregator} to build protected interest representations $\hat{\textbf{r}}_i$ and perform news recall to obtain candidate news set $\mathcal{R}$.
The recalled news with their titles and links are further distributed to the user client.
After the client receives the candidate news, the ranking model locally ranks these candidate news and displays the top $D$ news with the highest-ranking scores.
The user can locally read news titles and click links of interested news for reading.
Finally, after the user finishes this session, all displayed news and user behaviors will be stored in the user client.
Besides, to reduce online latency, protected decomposition weights and interest channel quotas can be calculated and stored in advance for quick uploading.

\subsection{Privacy-Preserving Model Training}

Training accurate recommendation models usually relies on large-scale training data~\cite{muhammad2020fedfast}.
In \textit{Uni-FedRec}, user behavior data is locally stored in user clients and it is hard to centrally learn parameters of recommendation models in a conventional way.
Motivated by~\citet{mcmahan2017communication}, we utilize federated learning techniques to train recall and ranking models on decentralized user data.
Next, we will introduce the training process of the recall model in detail.\footnote{Training process of the ranking model is similar to the recall model and its description is omitted.}

As shown in Fig.~\ref{fig.training}, the server is in charge of maintaining parameters $\Theta_r$ of the recall model and coordinating clients to collaboratively update $\Theta_r$ based on user data locally stored on them.
At the beginning of a federated updating round, each client has a copy of the current recall model.
Then the server will randomly select a part of users $\mathcal{U}$ to perform local gradient calculation.
Client of the selected user $u$ will calculate gradients $\frac{\partial \mathcal{L}_u^r}{\partial \Theta_r}$ for parameter updating based on her behavior data $\mathcal{B}_u$ and current model parameters $\Theta_r$.
Besides, gradients calculated from user behavior data may leak private user information~\cite{deepleakge}.
To better protect user privacy, following \citet{qi2020privacy}, we apply a local differential privacy (LDP) module to the clipped gradients: 
\begin{equation}
    \hat{\textbf{G}}_u = f_\theta(\frac{\partial \mathcal{L}_u^r}{\partial \Theta_r}) + n_g, \quad  n_g\sim La(0,\lambda_g),
\end{equation}
where $\hat{\textbf{G}}_u$ is the protected gradients, $f_\theta$ is a clipping function with the scale of $\theta$, $n_g$ is the Laplace noise and $\lambda_g$ is its intensity.
Next, the user client uploads the protected gradients $\hat{\textbf{G}}_u$ to the server.

After the server receives uploaded gradients from clients of users in $\mathcal{U}$, the server further aggregates these gradients for model updating:
\begin{equation}
    \overline{\textbf{G}} = \sum_{u\in \mathcal{U}} \beta_u \hat{\textbf{G}}_u, \ \ \beta_u = \frac{|\mathcal{B}_u|}{\sum_{v \in \mathcal{U}} |\mathcal{B}_v|},
\end{equation}
where $\overline{\textbf{G}}$ is the average gradients.
Then, parameters of the recall model is updated as: $\Theta_r = \Theta_r - \omega \overline{\textbf{G}}$, where $\omega$ is the learning rate.
Updated parameters are further distributed to all clients to update local model.
We will repeat this federated updating process until the recall model training converges.

\subsection{Analysis on Privacy Protection}

In this section, we will analyze the privacy protection ability of \textit{Uni-FedRec}.
First, in \textit{Uni-FedRec}, users' behavior data is locally stored in their clients and is never uploaded to the server, which can effectively alleviate users' privacy concerns and risks of large-scale privacy leakage~\cite{mcmahan2017communication}.
To train models and serve users, \textit{Uni-FedRec} only needs to upload model gradients and user interest representations to the server.
These intermediate variables usually contain much less private information than raw data according to data processing inequality~\cite{mcmahan2017communication,qi2020privacy}.
Besides, these variables are aggregated from multiple behaviors of a user, making it more difficult to infer a specific user behavior from them.
Second, we propose an interest decomposer-aggregator method to protect interest representation $\textbf{r}_i$.
Since protected interest representation $\hat{\textbf{r}}_i$ is aggregated from basic interest embeddings shared among users instead of user's clicked news, it is more difficult to infer a specific user's clicked news from $\hat{\textbf{r}}_i$ than $\textbf{r}_i$.
Besides, in this method, $\textbf{r}_i\in \mathbb{R}^d$ which belongs to a $d$-dimensional space $\mathbb{R}^d$ is projected into a $B$-dimensional space $\mathbb{R}^B$.
Since $B$ is much smaller than $d$ in our settings, $\hat{\textbf{r}}_i$ can lose much information on user privacy.
Third, we apply the LDP technique to protect both interest representations and gradients.
Based on the LDP theory\cite{choi2018guaranteeing}, in \textit{Uni-FedRec}, the privacy budget upper bounds of protected gradients and protected interest representations can achieve $\frac{2\theta}{\lambda_g}$ and $\frac{2\delta}{\lambda_I}$, respectively.
Since a smaller privacy budget means better privacy protection ability, \textit{Uni-FedRec} can achieve a trade-off between model accuracy and privacy protection by adjusting noise intensity.

\section{Experiment}
\subsection{Experimental Datasets and Settings}

\begin{table}[h]
\centering
\resizebox{0.48\textwidth}{!}{
\begin{tabular}{cccccc}
\hline
      & \# News   & \# Users & \# Clicks  &\#Impressions \\ \hline
\textit{MIND}  & 161,013  & 1,000,000   & 24,155,470 & 15,777,377   \\
\textit{NewsFeeds} & 120,219  & 20,000  & 112,927   & 48,923   \\ \hline
\end{tabular}
}
\caption{Dataset statistics.}
\label{table.stat}

\end{table}

\begin{table*}[]
\centering
\resizebox{0.99\textwidth}{!}{
\begin{tabular}{c|cccc|cccc}
\Xhline{1.5pt}
\multirow{2}{*}{} & \multicolumn{4}{c|}{MIND}     & \multicolumn{4}{c}{NewsFeeds} \\ \cline{2-9} 
                  & R@100 & R@200 & R@300 & R@400 & R@100  & R@200 & R@300 & R@400 \\ \hline
YoutubeNet               &1.50$\pm$0.03 &2.43$\pm$0.08 &3.34$\pm$0.08 &3.96$\pm$0.13 &0.60$\pm$0.02 &0.92$\pm$0.01 &1.17$\pm$0.01 &1.45$\pm$0.02\\
HUITA               &1.69$\pm$0.06 &2.67$\pm$0.04 &3.37$\pm$0.06 &3.97$\pm$0.06 &0.60$\pm$0.01 &0.91$\pm$0.01 &1.18$\pm$0.03 &1.45$\pm$0.01       \\
EBNR               &2.31$\pm$0.17 &3.72$\pm$0.13 &4.69$\pm$0.17 &5.61$\pm$0.17 &0.64$\pm$0.03 &0.96$\pm$0.05 &1.28$\pm$0.06 &1.55$\pm$0.06       \\
SASRec           &2.22$\pm$0.05 &3.51$\pm$0.07 &4.54$\pm$0.07 &5.38$\pm$0.07 &0.62$\pm$0.06 &0.96$\pm$0.01 &1.20$\pm$0.06 &1.49$\pm$0.05\\
PinnerSage        &1.22$\pm$0.14 &1.85$\pm$0.28 &2.69$\pm$0.23 &3.53$\pm$0.20 &0.59$\pm$0.01 &0.93$\pm$0.01 &1.15$\pm$0.01 &1.45$\pm$0.02 \\
Octopus           &1.26$\pm$0.03 &1.93$\pm$0.07 &2.74$\pm$0.06 &3.55$\pm$0.06 &0.60$\pm$0.02 &0.92$\pm$0.02 &1.17$\pm$0.02 &1.44$\pm$0.03\\\hline
Uni-FedRec          &\textbf{2.95}$\pm$0.11 &\textbf{4.13}$\pm$0.12 & \textbf{5.13}$\pm$0.12 &\textbf{5.99}$\pm$0.11&\textbf{0.80}$\pm$0.08 &\textbf{1.14}$\pm$0.10 & \textbf{1.60}$\pm$0.12 &\textbf{2.03}$\pm$0.12       \\ \Xhline{1.5pt}
\end{tabular}
}
\caption{News recall performance of different methods. Higher recall rates mean better performance. T-test on these results verifies the improvement of \textit{Uni-FedRec} over baseline methods is significant at level $p\le0.001$.
}
\label{tabel.recall}

\end{table*}

\begin{table*}[]
\centering
\resizebox{0.93\textwidth}{!}{
\begin{tabular}{c|cccc|cccc}
\Xhline{1.5pt}
\multirow{2}{*}{} & \multicolumn{4}{c|}{MIND}     & \multicolumn{4}{c}{NewsFeeds} \\ \cline{2-9} 
                  & R@100 & R@200 & R@300 & R@400 & R@100  & R@200 & R@300 & R@400 \\ \hline
YoutubeNet               &12.29 &15.91 &18.48 &20.64 &29.43 &31.22  &32.46 &33.47 \\
HUITA               &13.44  &16.11  &17.98  &19.49 &29.51  &31.24  &32.44  &33.39 \\
EBNR               &5.49  &8.27  &10.30  &12.05  &11.35  &13.08  &14.14 &14.86   \\
SASRec           &6.00  &8.71  &10.81  &12.52  &7.78  &9.18  &10.16  &11.03  \\
PinnerSage        &16.91  &21.35  &24.48 &27.18  &29.43  &31.10  &32.32  &33.38 \\
Octopus           &17.04  &21.62  &24.72  &27.31  &29.45  &31.15  &32.36  &33.38 \\\hline
Uni-FedRec          &\textbf{0.55} &\textbf{1.14}  & \textbf{1.69}  &\textbf{2.22} &\textbf{0.23}  &\textbf{0.54}  & \textbf{0.83}  &\textbf{1.08} \\ \Xhline{1.5pt}
\end{tabular}
}
\caption{Privacy protection performance of different methods, which is measured by rates of user's historical clicked news recalled from the news pool. Lower recall rates means better privacy protection performance.}
\label{tabel.privacy}
\end{table*}

We conduct experiments on two real-world datasets.
The first one is \textit{MIND}~\cite{wu2020mind}, a public news recommendation dataset constructed by logs of 1 million users in the Microsoft News during six weeks (Oct. 12 to Nov. 22, 2019).
The second one is \textit{NewsFeeds}, which is constructed by logs of 20,000 users from a commercial news feeds production of Microsoft during two weeks (Mar. 18 to Apri. 1, 2020).
User logs in the first week are used to construct historical user behaviors, logs in the last two days are used for evaluation, and other logs are used for model training.
More information on \textit{MIND} and \textit{NewsFeeds} is in Table~\ref{table.stat}.

Next, we will introduce settings of our experiments.
In our privacy-preserving news recall model, dimensions of both news and user representations are $256$.
The self-attention network contains $16$ attention heads with $16$-dimensional output vectors.
The clustering distance $d_c$ is set to 1.
The cluster-wise attention network is a two-layer dense network with $128$-dimensional hidden vector.
The number ($B$) of basic interest embeddings is set to $30$ and dimensions of these basic interest embeddings are 256.
The clipping scale $\delta$ is set to 0.2 and intensity $\lambda_I$ of the interest representation perturbation noise $n_I$ is set to 1.2.
Besides, we combine four different news ranking models, i.e., \textit{FedRec}~\cite{qi2020privacy}, \textit{NRMS}~\cite{wu2019neuralc}, \textit{LSTUR}~\cite{an2019neural} and \textit{NAML}~\cite{wu2019ijcai}, with our proposed privacy-preserving news recall model in \textit{Uni-FedRec}.
Embeddings generated by the news model and user model in these ranking methods are $256$-dimensional.
We randomly sample $r=2\%$ clients in each round for model updating.
The gradient clipping scale $\theta$ is $0.1$ and intensity $\lambda_g$ of gradient perturbation noise $n_g$ is $0.01$.
Negative sampling ratios for training both recall and ranking models, i.e., $K_r$ and $K_g$, are 4.
The learning rate $\omega$ is 0.05.
Codes of \textit{Uni-FedRec} are released for reproducing our method.\footnote{https://github.com/taoqi98/UniFedRec}

\subsection{News Recall Performance}
\label{Sec.privacy}

We compare the performance of different recall models on (1) news recall and (2) privacy protection.
News recall performance is measured by rates of users' future clicked news in the top $K$ recalled candidate news ($R@K$).
Privacy protection performance is measured by rates of users' historical clicked news in the top $K$ recalled candidate news.
Since it is easier to infer user private information from representations that can recall more users' historical clicked, methods that achieve lower recall rates on historical clicks are considered to have better privacy protection performance.
Here are baseline methods we compared:
(1) \textit{YoutubeNet}~\cite{covington2016deep}: averaging clicked news representations to recall candidate news.
(2) \textit{HUITA}~\cite{wu2019hierarchical}: attentively aggregating clicked news for news recall.
(3) \textit{EBNR}~\cite{okura2017embedding}: learning user representation via a GRU network.
(4) \textit{SASRec}~\cite{kang2018self}: using a self-attention network to learn user representation.
(5) \textit{PinnerSage}~\cite{pal2020pinnersage}: learning multiple interest representations via news clustering.
(6) \textit{Octopus}~\cite{liu2020octopus}: modding multiple user interest via elastic archive network.

We repeat experiment on each method $5$ times and show results in Table~\ref{tabel.recall} and Table~\ref{tabel.privacy}.
As shown in Table~\ref{tabel.recall}, \textit{Uni-FedRec} significantly outperforms baseline methods in news recall.
This is because users usually have diverse interests, and it is difficult for baselines to comprehensively model user interests.
Different from these methods, \textit{Uni-FedRec} learns multiple interest representations for a user from clusters of clicked news, which can comprehensively model diverse user interests in different fields.
Besides, as shown in Table~\ref{tabel.privacy}, \textit{Uni-FedRec} can better protect user privacy than baseline methods.
This is because baseline methods build user interest representations from the aggregation user's clicked news, making user's clicked news can be easily inferred from interest representations, which raises privacy leakage risks.
Different from these methods, we propose to synthesize interest representations by combing privacy-insensitive basic interest embeddings shared among different users instead of user's clicked news, which can better protect user privacy encoded in user representations.

\subsection{News Recommendation Performance}

\begin{table}[]
\centering
\resizebox{0.49\textwidth}{!}{
\begin{tabular}{c|cccc}
\Xhline{1.5pt}
          & {FedRec} & {LSTUR} & {NRMS} & {NAML} \\ \hline
{YoutubeNet}        &70.65 &68.53 &68.79 &65.93   \\
{HUITA}        &70.48 &68.76 &70.48 &68.76\\
{EBNR}       &75.56 &73.82 &75.01 &70.89\\
{SASRec}    &75.07 &72.51 &73.35 &70.51\\
{PinnerSage} &69.26 &68.96 &67.28 &66.09    \\
{Octopus}    &69.76 &69.12 &67.11 &65.75\\ \hline
{Uni-FedRec}  &\textbf{79.26} &\textbf{77.31} & \textbf{78.91} &\textbf{75.40}
 \\ \Xhline{1.5pt}
\end{tabular}
}
\caption{Recommendation performance (AUC) of different methods on \textit{MIND}, where rows and columns are different recall and ranking methods, respectively.}
\label{table.ranking}

\end{table}

Next, we combine different recall models and ranking models to evaluate the overall recommendation performance.
We first use the recall model to generate $400$ candidate news and further use the ranking model to rank these candidate news.
We use users' real click behaviors as ground truth and report AUC scores.
Experimental results are presented in Table~\ref{table.ranking}, and we only show results on \textit{MIND} dataset in the following sections due to space limitation.
Results show that \textit{Uni-FedRec} can consistently outperform baseline recall models when they are combined with a same ranking model.
These results further verify that \textit{Uni-FedRec} can outperform baseline methods in recommendation accuracy.

\subsection{Ablation Study}
\label{Sec.ablation}

\begin{figure}
    \centering
    \resizebox{0.49\textwidth}{!}{
    \includegraphics{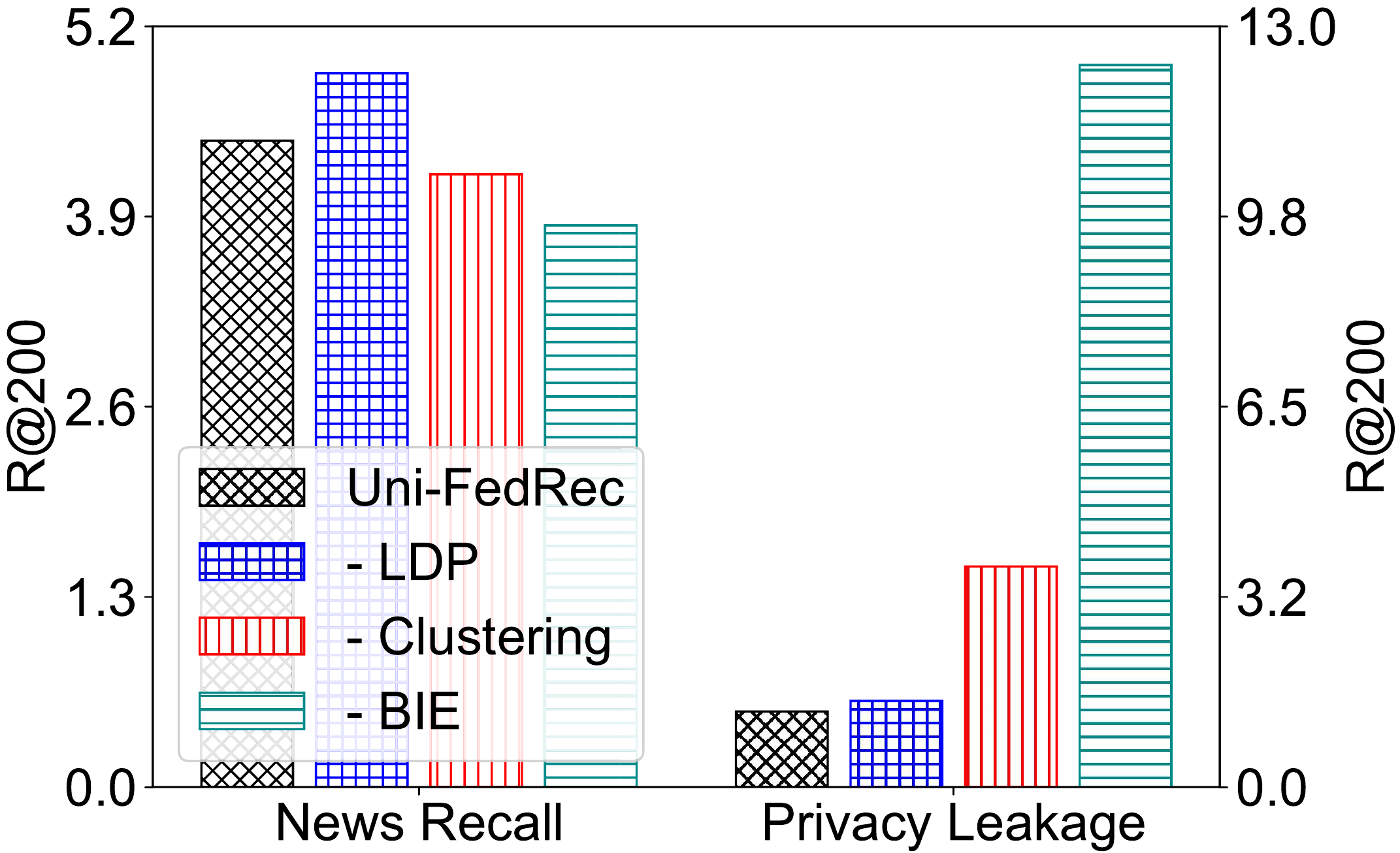}
    }
    \caption{Ablation study on \textit{Uni-FedRec}.}
    \label{fig.ablation}
\end{figure}

As shown in Fig.~\ref{fig.ablation}, we verify the effectiveness of important modules in \textit{Uni-FedRec} by removing them.
First, after removing the LDP module in the recall model, news recall performance of \textit{Uni-FedRec} improves while the privacy protection performance declines.
This is intuitive since perturbation noise will make \textit{Uni-FedRec} less accurate.
Second, removing the hierarchical clustering framework hurts the news recall performance.
This is because a user usually has diverse interests, which can be more comprehensively modeled by multiple interest representations.
Third, removing \textit{BIE} seriously hurts the privacy protection performance.
This is because protected interest representations are synthesized from basic interest embedding shared among different users, which contain much less private information of a specific user.

\subsection{Influence of the LDP Noise}

\begin{figure}
    \centering
    \resizebox{0.43\textwidth}{!}{
    \includegraphics{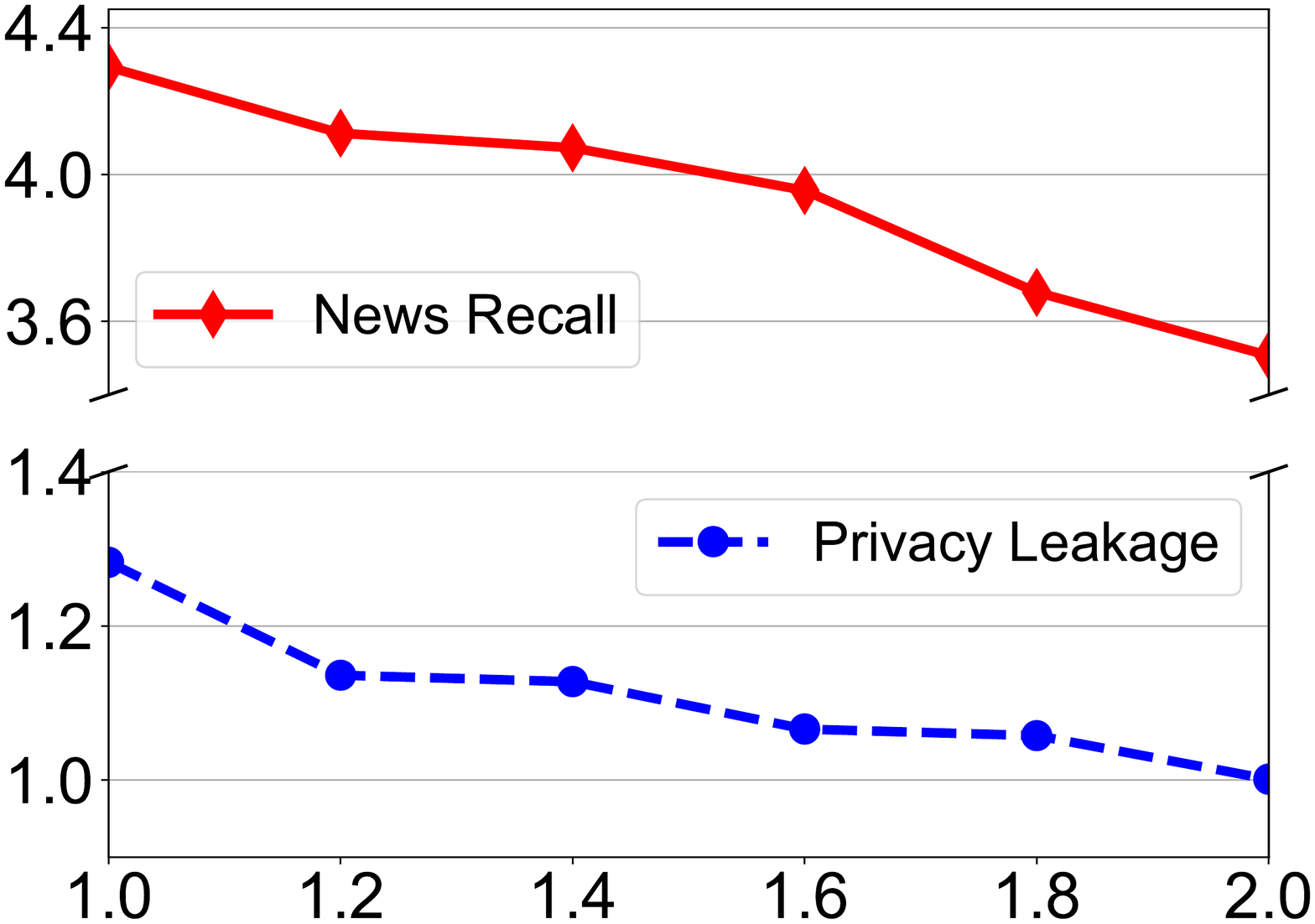}
    }
    \caption{Influence of $\lambda_I$ on \textit{Uni-FedRec}.}
    \label{fig.hyper.lambd}
\end{figure}

As shown in Fig.~\ref{fig.hyper.lambd}, we evaluate the influence of intensity $\lambda_I$ of LDP noise $n_I$ on \textit{Uni-FedRec}.
We find that with the increase of $\lambda_I$, news recall performance of \textit{Uni-FedRec} declines and the privacy protection ability of \textit{Uni-FedRec} increases.
This is intuitive since incorporating larger noise will more seriously hurt the information capability of interest representations on both user interests and user privacy.
Results in Fig.~\ref{fig.hyper.lambd} inspire us that we can find a trade-off between recommendation accuracy and privacy protection by adjusting the intensity $\lambda_I$ of LDP noise $n_I$ on interest representations.

\subsection{Influence of Clustering Distance}

\begin{figure}
    \centering
    \resizebox{0.43\textwidth}{!}{
    \includegraphics{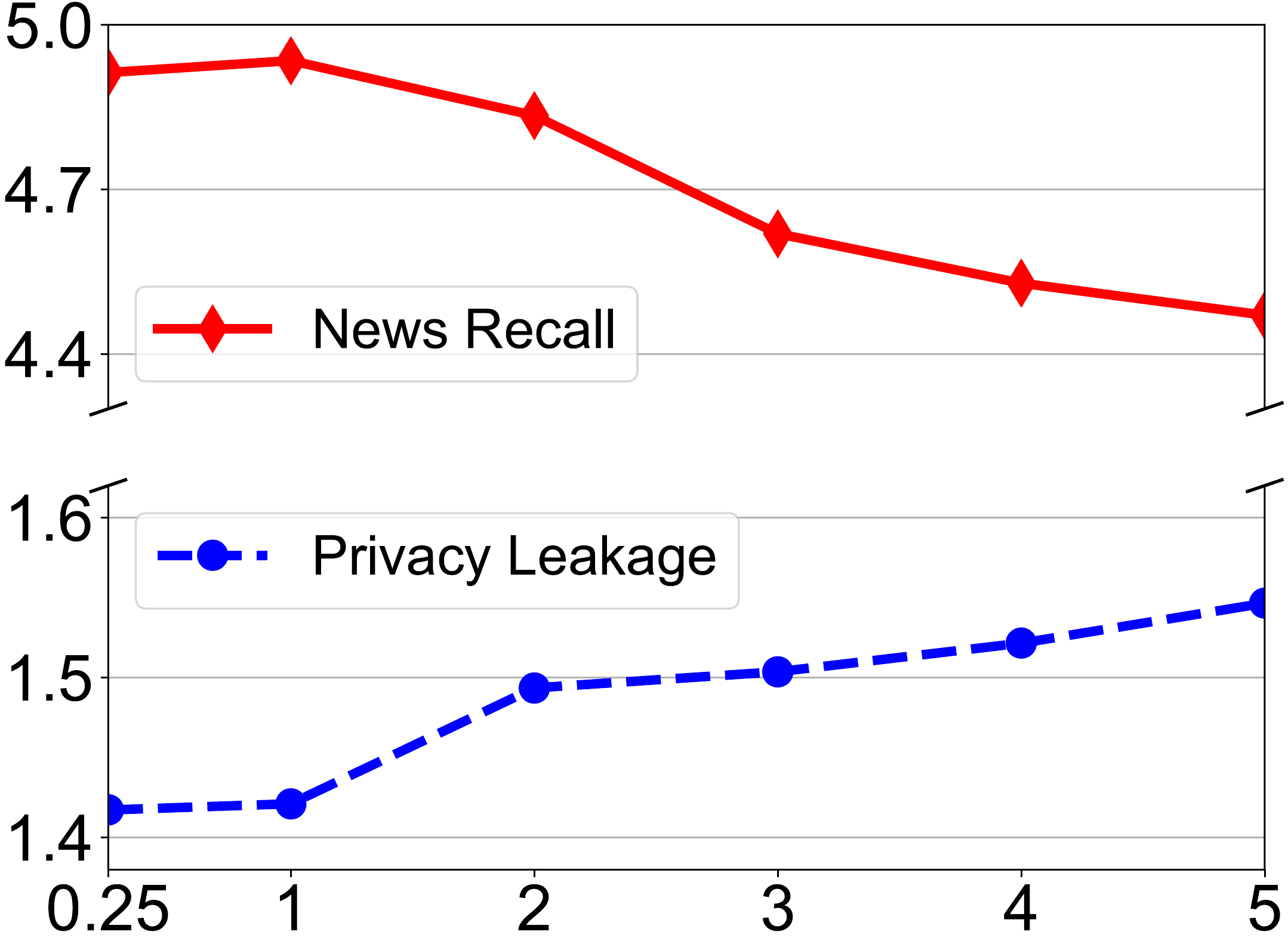}
    }
    \caption{Influence of $d_c$ on \textit{Uni-FedRec}.}
    \label{fig.hyper.distance}
\end{figure}

In Fig.~\ref{fig.hyper.distance}, we show the influence of clustering distance threshold $\textit{d}_c$ on \textit{Uni-FedRec}.
First, after $d_c$ increases, recall performance of \textit{Uni-FedRec} first increases.
This is because small $d_c$ makes \textit{Uni-FedRec} build too many interest clusters, which may bring some noise and hurt the accuracy of interest representations.
Second, when $d_c$ becomes large enough, recall performance begins to decline.
This is because larger $d_c$ makes \textit{Uni-FedRec} build fewer interest clusters and make it harder to comprehensively cover diverse user interests.
Third, with the increase of $d_c$, the privacy protection performance of \textit{Uni-FedRec} declines.
This may be because when \textit{Uni-FedRec} contains more interest channels, a single interest representation contains less private information.
It may be easier for our proposed interest decomposer-aggregator method to protect private information encoded in them.
Thus, a moderate value of $d_c$, i.e., 1, is suitable for \textit{Uni-FedRec}.

\subsection{ Training Convergence Analysis}

\begin{figure}
    \centering
    \resizebox{0.46\textwidth}{!}{
    \includegraphics{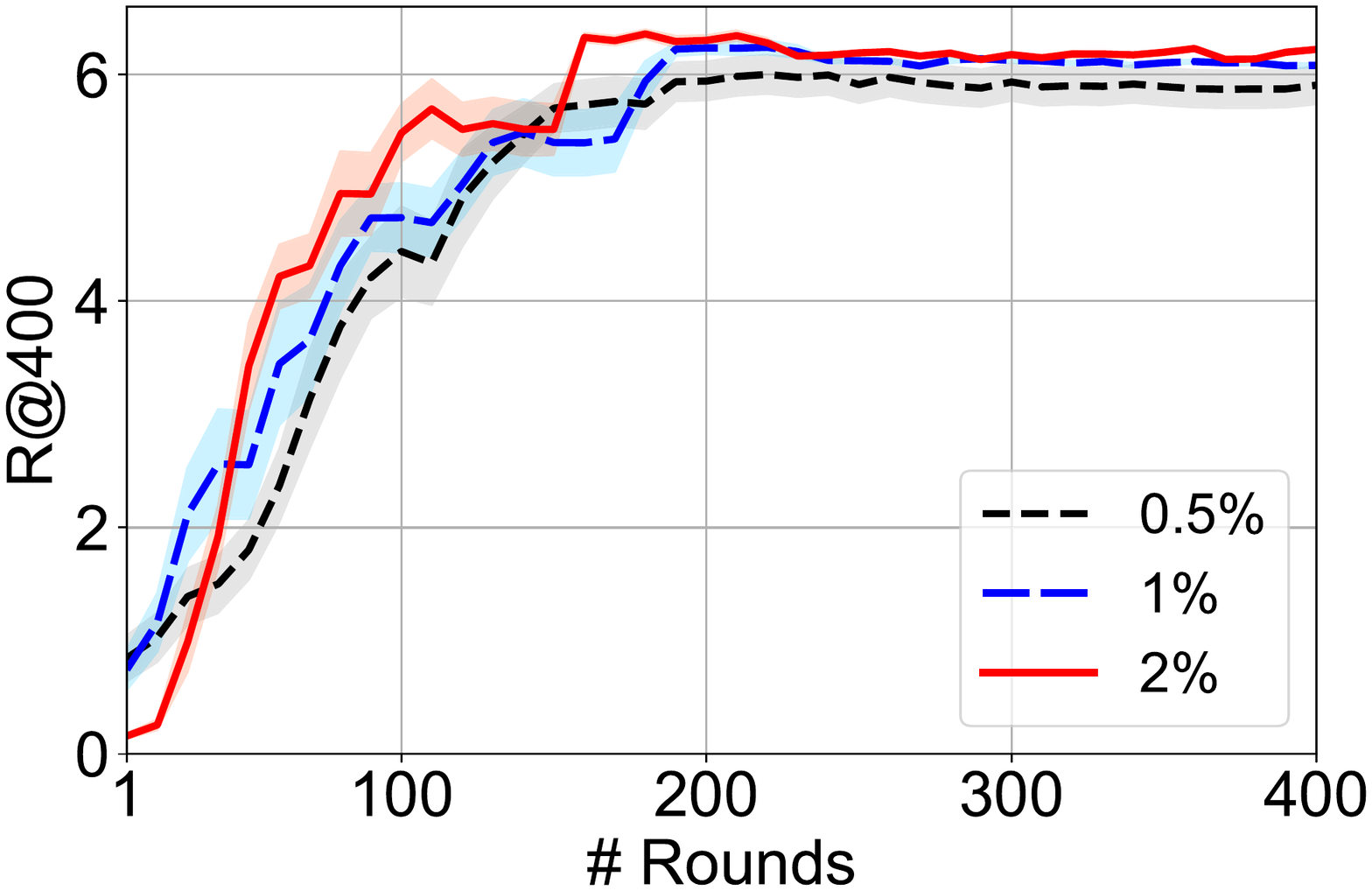}
    }
    \caption{Convergence curves in model training.}
    \label{fig.conv}
\end{figure}

Fig.~\ref{fig.conv} shows the model training convergence of our recall model.
The model is trained with different ratios of clients for a single federated model updating round.
First, training of \textit{Uni-FedRec} can usually converge in no more than two hundred steps, which verifies the efficiency of federated training of our recall model.
Second, training convergence of \textit{Uni-FedRec} will get faster and more stabilized if more clients can participate a single updating round.
This is intuitive since updating parameters with more training data is usually more accurate.

\section{Conclusion}

In this paper, we propose a unified privacy-preserving news recommendation framework (\textit{Uni-FedRec}) that can utilize user data locally stored in user clients to train models and serve users.
Our framework contains a recall stage and a ranking stage.
In the recall stage, the user client first employs a recall model to locally learn multiple interest representations from clicked news to model diverse user interest, which are further uploaded to the server to recall candidate news from a news pool.
In the ranking stage, candidate news are distributed to the user client and locally ranked for personalized display.
Besides, we propose an interest decomposer-aggregator method with permutation noise to protect private user information encoded in interest representations.
In addition, \textit{Uni-FedRec} collaboratively trains recall and ranking models on user data decentralized in massive user clients in a privacy-preserving way.
Experiments on two real-world datasets show that our method can significantly outperform baseline methods and effectively protect user privacy.

\section*{Acknowledgments}

This work was supported by the National Natural Science Foundation of China under Grant numbers U1936216, U1705261, and Tsinghua-Toyota Research Funds 20213930033.

\bibliography{main}
\bibliographystyle{acl_natbib}

\end{document}